# Generalized Uncertainty Principles for Quantum Cryptography

Randy Kuang
Quantropi Inc., Ottawa Canada
Email: randy.kuang@quantropi.com

*Abstract*—We know the classical public cryptographic algorithms are based on certain NP-hard problems such as the integer factoring in RSA and the discrete logarithm in Diffie-Hellman. They are going to be vulnerable with fault-tolerant quantum computers. We also know that the uncertainty principle for quantum bits or qubits such as quantum key distribution or QKD based on the quantum uncertainty principle offers the information theoretical security. The interesting implication with the paradigm shifts from classical computing to quantum computing is that the NP-hardness used for classical cryptography may shift to the uncertainty principles for quantum cryptography including quantum symmetric encryption, post-quantum cryptography, as well as quantum encryption in phase space for coherent optical communications. This paper would like to explore those so-called generalized uncertainty principles and explain what their implications are for quantum security. We identified three generalized uncertainty principles offering quantum security: non-commutability between permutation gates, non-commutability between the displacement and phase shift operators for coherent states, and the modular Diophantine Equation Problem in general linear algebra for post-quantum cryptography.

*Keywords—uncertainty principle, public key cryptography, post-quantum cryptography, quantum encryption, quantum security,*

I. INTRODUCTION

Today's information security is based on the foundation of the public key infrastructure or PKI with the well-known public key cryptographic algorithms such as RSA, Diffie-Hellman, elliptic curve Diffie-Hellman (ECDH) for key establishment and digital signature algorithm (DSA) or elliptic curve digital signature algorithm (ECDSA). The current PKI relies on the NP-hard problems such as the integer factoring problem in RSA and the discrete logarithm in Diffie-Hellman for key exchange and DSA or ECDSA for digital signature for information security within the scope of classical computing with the Boolean algebra. To break RSA-2048 public key requires a classical computer for 300 trillion years [1]. The Shor's algorithm provides a way to factor the RSA public key in polynomial time [2], with a fault-tolerant quantum computer. The current implementation by Skosana and Tame in 2021 [3] demonstrated the factorization for a number $21 = 3 \times 7$. The most recent preprint paper from a Chinese research team led by Professor Long has token a major step up to a new milestone to factor a 48-bit number 261980999226229 with a hybrid of classical and quantum approach [4]. Their algorithm is called Sublinear-resource Quantum Integer Factoring or SQIF. They achieved this milestone by leveraging both advantages from classical approach of the Schnorr's algorithm, turning a factoring problem into a lattice-based CVP problem [5,6] with a reduction from the Babai's algorithm for the optimal vector [7], and quantum approach, using a quantum optimizer called QAOA to replace the most time-consuming part of the Schnorr's algorithm for the so-called smooth-relation pairs. After obtaining enough smooth-relation pairs, the factors can be found by solving a linear equation system with a Boolean coefficient matrix. SQIF has two bounds B1 and B2 with B1 to be the quantum resource bound or the number of qubits which is a sublinear at $n = \mathcal{O}(m/\log_2 m)$ with $m$ to be the bit length of the factoring integer and B2 to be the number of smooth-relation pairs set to be at least $2n^2 + 1$. Comparing with the Shor's algorithm requiring 4099 logical qubits and billions of gate operations and circuit depth to break RSA-2048, SQIF requires only 372 good physical qubits for about 1200 circuit depth with the gate fidelity at level 99.9999%. Today, we do have quantum computers with more than 372 physical qubits, but their gate fidelity is about two-three nines. Once the gate fidelity is improved to the required level, RSA-2048 would be probably broken. From the capability perspective, SQIF looks more promising to reach that threshold of breaking RSA-2048 earlier than using the Shor's algorithm. On the other hand, the authors [4] suggested in a private conversation to make a trade-off with more physical qubits for the lower gate fidelity to increase the capability of factoring integers in today's or near future quantum computers. SQIF algorithm has been proven by Hegade and Enrique using digitized-counterdiabatic quantum computing or DCQC [8], replacing QAOA quantum optimizer. Hegade and Enrique also laid out their roadmap for factoring RSA-64, RSA-128, and RSA-2048 with a hybrid technique.

The fast advancement in quantum computing drives the urgency of quantum cryptography, especially quantum safe asymmetric cryptography or post-quantum cryptography called PQC. After three rounds of NIST standardization reviews, NIST announced its first standardized algorithms [9]: the lattice-based Kyber [10] for key encapsulation mechanism or KEM, the lattice-based Dillithium [11] and Falcon [12], and hash-based SPHINCS+ [13] for digital signature. Other KEM candidates BIKE [14], Classic McEliece [15], HQC [16], and SIKE [17] moved to the 4th round to be considered further.

NIST has also announced its reopening submissions for digital signature standardization due June 2023 [18].

Recent cryptanalysis reports have made two PQC candidates be vulnerable: Supersingular Isogeny Diffie-Hellman protocol or SIDH [19, 20] and Multivariate Public Key Cryptography or MPKC [21]. Another interesting research reported by Emily, et al. in 2022 demonstrated that the Short Vector Problem or SVP in lattice-based algorithms can be solved by using machine learning for lattice dimension from small to mid size [22]. Plus, the SQIF paper also demonstrates the capability to solve SVP problem with reduced quantum resources [4]. As continuing advancement of quantum computing, we would face more and more cryptanalyses from quantum inspired and quantum computing-aided algorithms as what we have seen since 2022. This trend of cryptanalysis should wake us up to rethink what the problem quantum cryptography should root on to defend the incoming powerful quantum computing threat.

Looking back to One-Time-Pad or OTP proven to be perfect secrecy by Shannon in 1949 [23]. Although OTP is a perfectly secure encryption with the true random key of the same length as the plaintext, it can only be used for once because the encryption is the Boolean logic operator XOR which is commutative and self-cancelled out if the same key is reused for encryption. On the other hand, quantum key distribution or QKD, proposed by Bennett and Brassard in 1984 [24], has been proven to be information theoretical secure by Shor and Preskill in 2000 [25]. If we look at OTP and QKD more closely, both encryption schemes are associated with each other once we describe them in the quantum computational basis $\{|0\rangle, |1\rangle\}$ where XOR is expressed with a matrix $XOR = \begin{bmatrix} 0 & 1 \\ 1 & 0 \end{bmatrix}$ for OTP and QKD encoding is expressed with Hadamard gate $H = \frac{1}{\sqrt{2}}\begin{bmatrix} 1 & 1 \\ 1 & -1 \end{bmatrix}$. The magic here is that the Hadamard gate is nothing else but the eigen-basis of XOR operator. Quantum mechanically, the computational basis denoted by $\{|0\rangle, |1\rangle\}$ for a single qubit is the eigen-basis of information operator $\hat{I}$ with two eigenvalues $I_1$ referring to information bit value 0 and $I_2$ referring to information bit value 1. Then the information operator $\hat{I}$ and XOR operator $\widehat{XOR}$ have their matrix expressions $\hat{I} = \begin{bmatrix} I_1 & 0 \\ 0 & I_2 \end{bmatrix}$ and $\widehat{XOR} = \begin{bmatrix} 0 & 1 \\ 1 & 0 \end{bmatrix}$. What we know the uncertainty principle in QKD could be specifically expressed as follows

$$\hat{I}\,\widehat{XOR} - \widehat{XOR}\,\hat{I} = \begin{bmatrix} I_1 & 0 \\ 0 & I_2 \end{bmatrix}\begin{bmatrix} 0 & 1 \\ 1 & 0 \end{bmatrix} - \begin{bmatrix} 0 & 1 \\ 1 & 0 \end{bmatrix}\begin{bmatrix} I_1 & 0 \\ 0 & I_2 \end{bmatrix} \neq 0.$$

Therefore, the information theoretical security of QKD proven by Shor and Preskill in 2000 [25] has its quantum mechanical interpretation: a generalized uncertainty principle between operators $\hat{I}$ and $\widehat{XOR}$ implemented physically with qubits. Theoretically, we can continue to extend QKD encoding mechanism for two or more qubits by leveraging the quantum uncertainty principle for multiple qubit system and quantum superpositions. However, it is very impractical to implement it that way, even just for two qubit QKD.

The objective of this paper is to describe and demonstrate those generalized uncertainty principles identified in our research for quantum symmetric and asymmetric cryptographic encryptions, as well as quantum encryption in phase space for coherent communications.

## II. SECURITY AND GENERALIZED UNCERTAINTY

### A. Shannon Perfect Secrecy in Boolean Algebra

Shannon stated in 1949 [23] that ***Perfect systems in which the number of cryptograms, the number of messages, and the number of keys are all equal are characterized by the properties that (1) each is connected to each by exactly one line, (2) all keys are equally likely. Thus the matrix representation of the system is a "Latin square"***. Let's use a toy example of items $0, 1, 2, 3$ to demonstrate the Shannon perfect cryptosystem. The Latin square is written as follows:

$$\begin{bmatrix} 0 & 1 & 2 & 3 \\ 1 & 0 & 3 & 2 \\ 2 & 3 & 0 & 1 \\ 3 & 2 & 1 & 0 \end{bmatrix}$$

which denotes the plaintext space, the ciphertext space and the key space are identical: $\{0, 1, 2, 3\}$. This Latin square can be decomposed into following 4 matrix vector operations:

$$\begin{bmatrix} 1 & 0 & 0 & 0 \\ 0 & 1 & 0 & 0 \\ 0 & 0 & 1 & 0 \\ 0 & 0 & 0 & 1 \end{bmatrix}\begin{bmatrix} 0 \\ 1 \\ 2 \\ 3 \end{bmatrix} = \begin{bmatrix} 0 \\ 1 \\ 2 \\ 3 \end{bmatrix}, \quad \begin{bmatrix} 0 & 1 & 0 & 0 \\ 1 & 0 & 0 & 0 \\ 0 & 0 & 0 & 1 \\ 0 & 0 & 1 & 0 \end{bmatrix}\begin{bmatrix} 0 \\ 1 \\ 2 \\ 3 \end{bmatrix} = \begin{bmatrix} 1 \\ 0 \\ 3 \\ 2 \end{bmatrix}$$

$$\begin{bmatrix} 0 & 0 & 1 & 0 \\ 0 & 0 & 0 & 1 \\ 1 & 0 & 0 & 0 \\ 0 & 1 & 0 & 0 \end{bmatrix}\begin{bmatrix} 0 \\ 1 \\ 2 \\ 3 \end{bmatrix} = \begin{bmatrix} 2 \\ 3 \\ 0 \\ 1 \end{bmatrix}, \quad \begin{bmatrix} 0 & 0 & 0 & 1 \\ 0 & 0 & 1 & 0 \\ 0 & 1 & 0 & 0 \\ 1 & 0 & 0 & 0 \end{bmatrix}\begin{bmatrix} 0 \\ 1 \\ 2 \\ 3 \end{bmatrix} = \begin{bmatrix} 3 \\ 2 \\ 1 \\ 0 \end{bmatrix} \quad (1)$$

denoting 4 encryptions in terms of matrix and vector multiplications $\widehat{K}_i\,\vec{m} = \vec{c}_i$ with $i = 0, 1, 2, 3$. The square matrix on the left of an equation represents the encryption transformation with a particular key taking from the key space $\{0, 1, 2, 3\}$. All transformation matrices are unitary and symmetric and commutative. It can be easily verified that $c_{ij} = i \oplus j$. That means, the Shannon perfect cryptosystem performs encryption and decryption with Boolean XOR operator. Then the perfect secrecy of the Shannon cryptosystem is solely dependent on the key randomly chosen from the key space or the uncertainty of the key space.

### B. Extension of Shannon Perfect Secrecy to Linear Algebra

Kuang and Bettenburg in 2020 [26] have extended the Shannon perfect cryptosystems, representing with the Latin square in Boolean algebra, to a generic n-bit or qubit system by replacing the XOR symmetric permutation matrices with any n-bit permutation matrices. There are total $(2^n!)$ permutation matrices from the quantum computational basis $\{|0\rangle, |1\rangle, \dots, |2^n - 1\rangle\}$. Of those $(2^n!)$ permutation matrices, there are $2^n$ matrices representing XOR operations between n-bit numbers as what we have seen in the Shannon perfect cryptosystem and other $(2^n - 1)!$ permutation matrices cannot be simply expressed in terms of any known Boolean operators which leads to another uncertainty of choosing encryption operators. With this extension, the Shannon perfect systems are then characterized by the properties that *(1) each is connected to each by exactly* $(2^n − 1)!$ *lines, (2) all keys (permutation matrices) are equally likely [26].*

Kuang and Bettenburg proposed to map the classical key into a set of permutation matrices called Quantum Permutation Pad or QPP, using some algorithm such as the Fisher-Yates random shuffling. Kuang and Barbeau in 2022 proposed a scheme of universal cryptography using QPP [27] and a few use cases were implemented in classical computing systems [28, 29, 30, 31, 32]. Kuang and Perepechaenko in 2022 reported their implementations in IBM Quantum computers by directly using the permutation gates for permutation matrices to create quantum permutation circuits [33, 34, 35]. These toy examples demonstrated that QPP can be implemented in both classical and quantum native systems so QPP could be used to bridge quantum encrypted communications between quantum and quantum, classical and classical, and quantum and classical systems.

### III. GENERALIZED UNCERTAINTY PRINCIPLES

The uncertainty principle in quantum physics is the fundamental principle which states that two non-commutative operators cannot be accurately measured at the same time in any same basis. QKD turns this general uncertainty principle $\hat{I}\,\widehat{XOR} - \widehat{XOR}\,\hat{I} \neq 0$ into secure key establishment by encoding an information bit in a randomly selected basis of two conjugate bases of the information operator $\hat{I}$ and XOR operator $\widehat{XOR}$. Encoding in the eigen-basis of $\widehat{XOR}$ is equivalent to the Hadamard gate operation. That means, the uncertainty principle in QKD leads to the physical untouchable security, in contrast to the uninterpretable security of the Shannon perfect cryptosystem. It may be wisely to have a cryptography or quantum cryptography built on both **the uncertainty** of choosing encryption operators from a large key (operator) space and **the uncertainty** of non-commutativity between chosen operators which offers both perfect in encryption and reusable with the chosen operators. That is what we refer to as the generalized uncertainty principles for quantum cryptography.

We identified three generalized uncertainty principles for quantum cryptography in symmetric and asymmetric encryption, as well as quantum analogue encryption for coherent optical communications.

#### A. Uncertainty within Permutation Gates

In a general case of n-qubit system, there are $(2^n!)$ permutation gates existing over the computational basis $\{|0\rangle, |1\rangle, ..., |2^n − 1\rangle\}$ which form a quantum permutation space or key operator space.

**The first Generalized Uncertainty Principle**: Under the computational basis $\{|0\rangle, |1\rangle, ..., |2^n − 1\rangle\}$ of n-qubit system, there exists a general non-commutativity between two quantum permutation gates $\hat{p}_i$ and $\hat{p}_j$

$$\hat{p}_i\hat{p}_j - \hat{p}_j\hat{p}_i \neq 0, \quad \text{for } i \neq j = 1, 2, ..., 2^n! \tag{2}$$

The entire space of quantum permutation gates holds the uncertainty $(2^n!)$ of the chosen permutation gates and the uncertainty between permutation gates as shown in Eq. (2) leading to the reusability of the chosen encryption gates. Therefore, this generalized uncertainty principle leads the generalized Shannon perfect cryptosystem [26, 27]. For details of how QPP can be used to perform encryption and decryption, please refer to the publications [26, 27, 28, 29, 30, 31, 32] for classical computing system and [33, 34, 35] for quantum native computing system. Here we just illustrate the general scenario of implementation in classical systems in Fig. 1.

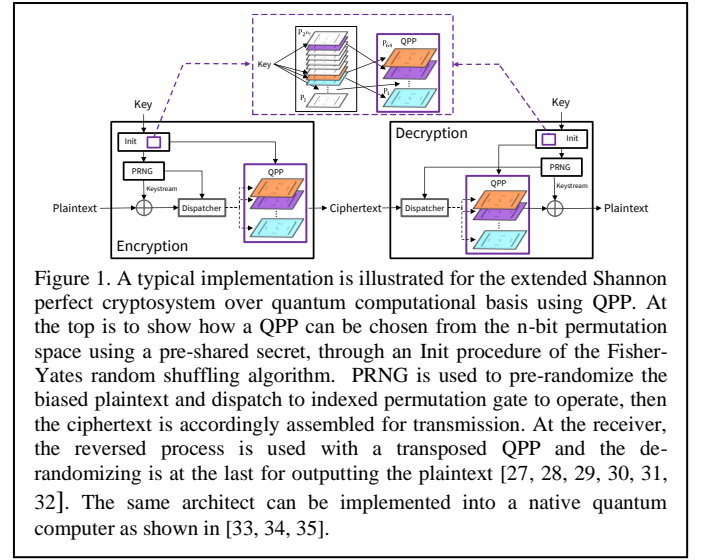

Figure 1. A typical implementation is illustrated for the extended Shannon perfect cryptosystem over quantum computational basis using QPP. At the top is to show how a QPP can be chosen from the n-bit permutation space using a pre-shared secret, through an Init procedure of the Fisher-Yates random shuffling algorithm. PRNG is used to pre-randomize the biased plaintext and dispatch to indexed permutation gate to operate, then the ciphertext is accordingly assembled for transmission. At the receiver, the reversed process is used with a transposed QPP and the de-randomizing is at the last for outputting the plaintext [27, 28, 29, 30, 31, 32]. The same architect can be implemented into a native quantum computer as shown in [33, 34, 35].

#### B. Uncertainty between Displacement and Phase Shift Operators

For the coherent states $|\beta\rangle$ of a quantum harmonic oscillator, the displacement operator $\widehat{D}(\alpha)$ is defined as

$$\widehat{D}(\alpha) = e^{\alpha\,\hat{a}^\dagger - \alpha^*\,\hat{a}} \tag{3}$$

and it is easy to verify that the displacement operator is unitary and reversible: $\widehat{D}^{-1}(\alpha) = \widehat{D}^\dagger(\alpha) = \widehat{D}(-\alpha)$. Let's apply the displacement operator $\widehat{D}(\alpha)$ to a coherent state $|\beta\rangle$

$$\widehat{D}(\alpha)\,|\beta\rangle = \widehat{D}(\alpha)\,\widehat{D}(\beta)\,|0\rangle = e^{\alpha\beta^* - \alpha^*\beta}|\alpha + \beta\rangle \tag{4}$$

indicating that the displacement operator performs a transformation of the plain coherent state $|\beta\rangle$ into a cipher coherent state which is equivalent to an addition of two complex numbers $|\alpha + \beta\rangle$, except for a global phase which does not impact any physical measuring results. On the other hand, the phase shift operator is defined as

$$\hat{\varphi}(\phi) = e^{j\phi\,\hat{a}^\dagger\hat{a}} \tag{5}$$

and operating on a coherent state would change its phase angle,

$$\hat{\varphi}(\phi)|\beta\rangle = |e^{j\phi}\beta\rangle. \tag{6}$$

The phase shift operator is also unitary and reversible: $\hat{\varphi}^{-1}(\phi) = \hat{\varphi}^{\dagger}(\phi) = \hat{\varphi}(-\phi)$. Then we can easily verify the non-commutativity between the displacement and phase shift operators

$$\hat{D}(\alpha)\hat{\varphi}(\phi) - \hat{\varphi}(\phi)\hat{D}(\alpha) \neq 0 \quad (7)$$

**The second Generalized Uncertainty Principle**: In the phase space, there exists a generalized uncertainty principle between the displacement and phase shift operators $\hat{D}(\alpha)$ and $\hat{\varphi}(\phi)$

$$\hat{D}(\alpha)\hat{\varphi}(\phi) - \hat{\varphi}(\phi)\hat{D}(\alpha) \neq 0 \quad (8)$$

With $\alpha$ to be a complex number and $\phi \in [0, 2\pi)$. Eq. (8) is true for any given complex number $\alpha$ and phase shift $\phi$. Therefore, the uncertainty of chosen encryption operators $\hat{D}(\alpha)$ and $\hat{\varphi}(\phi)$ is large and they are also non-commutative.

Based on the second generalized uncertainty principle, an analogue encryption mechanism can be developed for coherent optical communications. Kuang and Bettenburg in 2020 proposed quantum public key envelope or QPKE with randomized phase shift [36], in a roundtrip scheme using self-shared random key mapped to phase shift for phase shift keying data modulation. Experimental implementation together with security analysis have been carried out since then [37, 38, 39, 40]. The experiments achieved the key distribution speed over 200 gbps at 80 km between Alice and

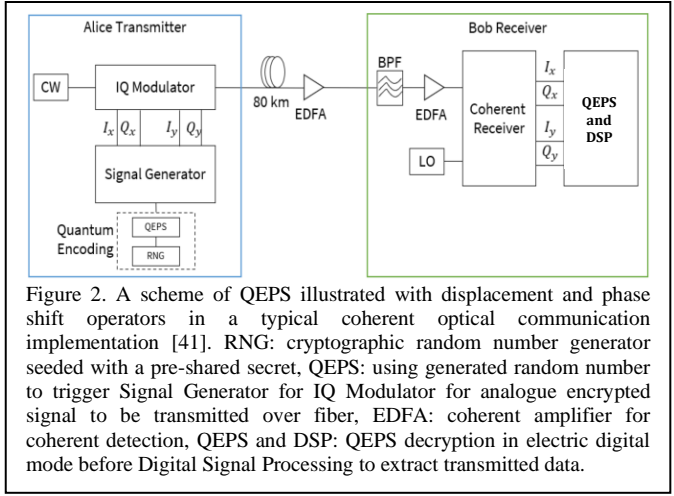

Figure 2. A scheme of QEPS illustrated with displacement and phase shift operators in a typical coherent optical communication implementation [41]. RNG: cryptographic random number generator seeded with a pre-shared secret, QEPS: using generated random number to trigger Signal Generator for IQ Modulator for analogue encrypted signal to be transmitted over fiber, EDFA: coherent amplifier for coherent detection, QEPS and DSP: QEPS decryption in electric digital mode before Digital Signal Processing to extract transmitted data.

Bob [39, 40]. Recently, Kuang and Chan proposed a Quantum Encryption in Phase Space or QEPS using the displacement operator together with phase shift operator or encryption operator $\hat{k}(\alpha, \phi) = \hat{D}(\alpha)\hat{\varphi}(\phi)$ with quadrature amplitude modulation or QAM and QPSK data modulations [41, 42]. Kuang and Chan [41] offered a security analysis using the wired-tap channel scheme. Due to their bijective transformation of $\hat{D}(\alpha)$ and $\hat{\varphi}(\phi)$, QEPS encryption can be also considered as the generalized Shannon perfect cryptosystem in an analogue domain.

Using QEPS for coherent encryption requires the communication peers to have a pre-shared secret to seed a cryptographic pseudo random number generator for synchronized random number generations. The random numbers from RNG are mapped to complex numbers $\alpha$ and phase shift $\phi$ for the displacement operator $\hat{D}(\alpha)$ and $\hat{\varphi}(\phi)$ at transmission side and $\hat{D}(-\alpha)$ and $\hat{\varphi}(-\phi)$ at receiving side. A simulation layout of QEPS with encryptor $\hat{k}(\alpha, \phi) = \hat{D}(\alpha)\hat{\varphi}(\phi)$ is illustrated in Fig. 2 and the simulation results are illustrated in Fig. 3 [41]. The direct detection of QEPS encrypted coherent states demonstrate a random constellation diagram as shown in Fig. 2(a) and 2(c). For any wiretapping with DSP processing or the Eve, the constellation diagram still shows a random scheme with a Bit-Error-Rate or BER = 47% in Fig. 2(c). The constellation in Fig. 2(b) shows a square-like random image but the DSP processing reveals a clear 16-QAM constellation with BER=0 as shown in Fig. 2(d).

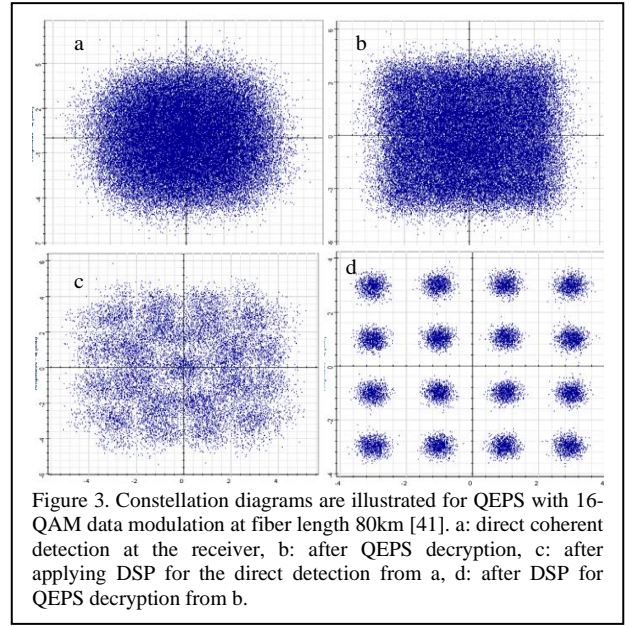

Figure 3. Constellation diagrams are illustrated for QEPS with 16-QAM data modulation at fiber length 80km [41]. a: direct coherent detection at the receiver, b: after QEPS decryption, c: after applying DSP for the direct detection from a, d: after DSP for QEPS decryption from b.

### C. Uncertainty from Modular Diophantine Equation Problem

In linear algebra, there are two special vector spaces: multidimensional vector space labeled with $k_1, k_2, ..., k_m$ and polynomial vector space labeled with $x^0, x^1, x^2, ..., x^N$ for a polynomial of order N. The variable $x \in \mathbb{F}_p$ in the polynomial is used to represent a message with $p$ to be a prime number and variables $k_1, k_2, ..., k_m \in \mathbb{F}_p$ are used to denote the random variables to be chosen at the time of encryption. The following special modular Diophantine Equation Problem or MDEP

$$p_1(x)k_1 + p_2(x)k_2 + \cdots + p_m(x)k_m = C \bmod p \quad (9)$$

where $p_i(x)$ are known polynomials of order N. Symbolically, a solution of the message $x$ can be expressed in terms of $k_1, k_2, ..., k_m$: $x = \mathcal{K}(k_1, k_2, ..., k_m) \in \mathbb{F}_p$. Therefore, the solution for $x$ is non-deterministic with a complexity $\mathcal{O}(p)$ of purely random guessing and a complexity $\mathcal{O}(p^m)$ of mathematically attacking. That means, the modular Diophantine Equation Problem is NP-complete as stated by Moore and Mertens in 2011 [43].

MDEP in Eq. (9) represents a map between the ciphertext $C$ and the plaintext $x$ with variables $k_1, k_2, ..., k_m \in \mathbb{F}_p$ as random encryption key. For each pair of $x$ and $C$, there generally exist $p^{m-1}$ set of solutions $(k_1, k_2, ..., k_m)$ met Eq. (13). This is another generalized Shannon perfect cryptosystem: *the Shannon perfect systems are then characterized by the properties that (1) each is connected to each by exactly $p^{m-1}$ lines, (2) all keys $(k_2, ..., k_m \in \mathbb{F}_p)$ are equally likely*. That leads to our next generalized uncertainty principle.

**The third Generalized Uncertainty Principle**: For a generic multivariate polynomial $G(x, k_1, k_2, ..., k_m)$ over a prime field $\mathbb{F}_p$ or a ring $\mathbb{Z}_p$, the well-known Modular Diophantine Equation Problem or MDEP can be written as

$$G(x, k_1, k_2, ..., k_m) = C \bmod p \quad (14)$$

with $x$ as a message variable, $k_1, k_2, ..., k_m$ as runtime encryption keys, and $C$ as its corresponding ciphertext of $x$ under the encryption with $k_1, k_2, ..., k_m$. The overall uncertainty of the keys is $p^m$, exponential complexity $\mathcal{O}(p^m)$ by attacking this cryptosystem in comparison with $\mathcal{O}(p)$ by purely random-guessing $x$. This generalized uncertainty principle can be used to develop algorithms for Post-Quantum Cryptography or PQC in KEM and DS.

Kuang, Perepechaenko, and Barbeau in 2022 proposed their Multivariate Polynomial Public Key or MPPK [44], relying on this generalized uncertainty principle against the secret recovery attack. Kuang and Perepechaenko further improved the security of MPPK for private key recovery attack by introducing homomorphic encryption on plain public key over hidden ring(s) then renaming MPPK as HPPK [45, 46]. We summarize the HPPK KEM [45] in Fig. 4 for illustration purpose. We recently identified the private key security of MPPK digital signature proposed by Kuang, Perepechaenko, and Barbeau in 2022 [47], relying on this generalized uncertainty principle too.

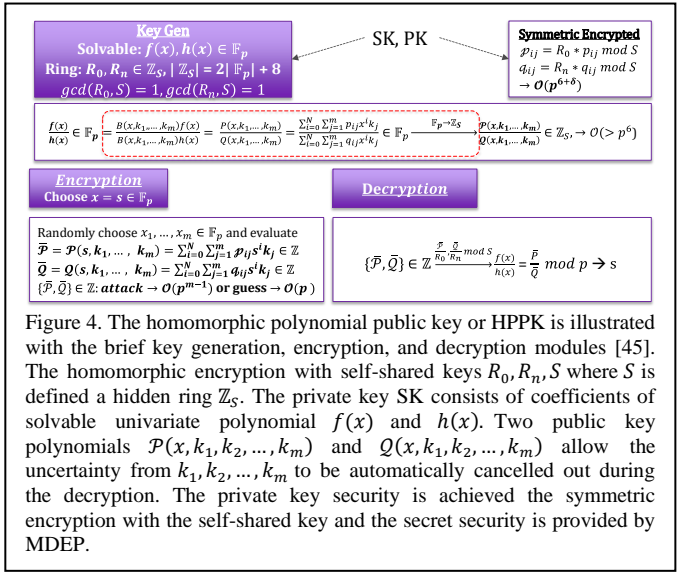

Figure 4. The homomorphic polynomial public key or HPPK is illustrated with the brief key generation, encryption, and decryption modules [45]. The homomorphic encryption with self-shared keys $R_0, R_n, S$ where $S$ is defined a hidden ring $\mathbb{Z}_S$. The private key SK consists of coefficients of solvable univariate polynomial $f(x)$ and $h(x)$. Two public key polynomials $\mathcal{P}(x, k_1, k_2, ..., k_m)$ and $\mathcal{Q}(x, k_1, k_2, ..., k_m)$ allow the uncertainty from $k_1, k_2, ..., k_m$ to be automatically cancelled out during the decryption. The private key security is achieved the symmetric encryption with the self-shared key and the secret security is provided by MDEP.

IV. DISCUSSION AND CONCLUSION

We introduced three generalized uncertainty principles based on the uncertainty of key space and the non-commutativity between operators in the key space. We summarize them in Fig.5. The well-known Shannon perfect cryptosystem in Boolean algebra at the top of Fig. 5 indicates a 1-to-1 mapping scheme with the Boolean XOR operator. The first generalized uncertainty principle refers to the QPP with a mapping $1 \xleftrightarrow{(2^n-1)!} 1$ for each pair of n-bit word plaintext and ciphertext. The QPP encryption scheme can be implemented in both classical system with matrix-vector multiplication and quantum system with permutation gates or circuits. The second generalized uncertainty principle can be applied to build KEM [45, 46] and DS [47] for PQC. The third generalized uncertainty principle helps to establish analogue quantum encryption in phase space such as QEPS for coherent optical communications or secure communications between future photonic quantum computers [41, 42].

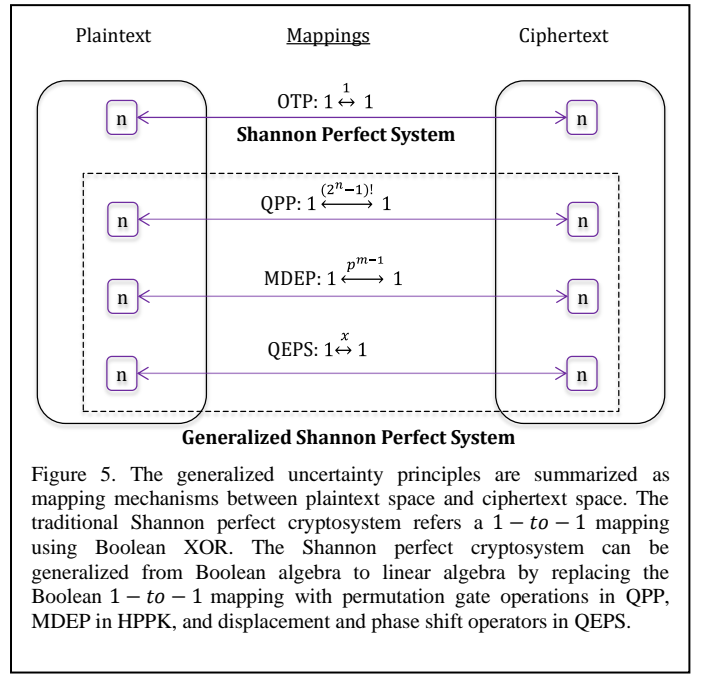

Figure 5. The generalized uncertainty principles are summarized as mapping mechanisms between plaintext space and ciphertext space. The traditional Shannon perfect cryptosystem refers a $1 - to - 1$ mapping using Boolean XOR. The Shannon perfect cryptosystem can be generalized from Boolean algebra to linear algebra by replacing the Boolean $1 - to - 1$ mapping with permutation gate operations in QPP, MDEP in HPPK, and displacement and phase shift operators in QEPS.

ACKNOWLEDGMENT

The author wishes to acknowledge Prof. Michel Barbeau from Carleton University, my colleagues Maria Perepechaenko and Adrian Chan for collaborations during the developments of quantum symmetric encryptions for universal gate quantum computers, multivariate polynomial public key schemes, and quantum encryption in phase space. The author also acknowledges M. Khalil, K. A. Shahriar, Prof. L. R. Chen, and Prof. D. V. Plant from McGill University for collaborations of experimental implementations of QEPS with the phase shift operator.